\documentclass[12pt]{article}
\usepackage{graphicx}

      \usepackage[cp1251]{inputenc} 

\begin{document}

{\noindent \small Astronomy Letters, 2012, Vol. 38, No. 10, pp.
638-648}

\medskip

      \bigskip
      \centerline {\large\bf Estimation of the Galactic Spiral Pattern Speed from Cepheids}

      \bigskip
      \centerline {V. V. Bobylev and A. T. Bajkova}

      \bigskip
 \centerline {\small \it Pulkovo Astronomical
Observatory, Russian Academy of Sciences, }
 \centerline {\small \it Pulkovskoe sh. 65, St.
Petersburg, 196140 Russia}

\medskip

{\noindent To study the peculiarities of the Galactic spiral
density wave, we have analyzed the space velocities of Galactic
Cepheids with proper motions from the Hipparcos catalog and
line-of-sight velocities from various sources. First, based on the
entire sample of 185 stars and taking $R_0 = 8$ kpc, we have found
the components of the peculiar solar velocity
$(u_\odot,v_\odot,w_\odot)=(7.6,11.6,6.1)\pm(0.8,1.1,0.6)$ km
s$^{-1}$, the angular velocity of Galactic rotation $\Omega_0 =
-27.4\pm0.6$ km s$^{-1}$ kpc$^{-1}$ and its derivatives
$\Omega^{'}_0 = +4.07\pm0.21,$ km s$^{-1}$ kpc$^{-2}$ and
$\Omega^{''}_0 = -0.83\pm0.17,$ km s$^{-1}$ kpc$^{-3}$, the
amplitudes of the velocity perturbations in the spiral density
wave $f_R=-6.7\pm0.7$ and $f_\theta= 3.5\pm0.5$ km s$^{-1}$, the
pitch angle of a two-armed spiral pattern (m = 2)
$i=-4.5\pm0.1^\circ$ (which corresponds to a wavelength
$\lambda=2.0\pm0.1$ kpc), and the phase of the Sun in the spiral
density wave $\chi_\odot=-191\pm5^\circ$. The phase $\chi_\odot$
has been found to change noticeably with the mean age of the
sample. Having analyzed these phase shifts, we have determined the
mean value of the angular velocity difference $\Omega_p-\Omega$,
which depends significantly on the calibrations used to estimate
the individual ages of Cepheids. When estimating the ages of
Cepheids based on Efremov's calibration, we have found
$|\Omega_p-\Omega_0|=9\pm2$ km s$^{-1}$ kpc$^{-1}$. The ratio of
the radial component of the gravitational force produced by the
spiral arms to the total gravitational force of the Galaxy has
been estimated to be $f_{r0} = 0.04$.}

     \bigskip\noindent
      {\it Keywords:} Cepheids, spiral structure, Galactic
      kinematics.

      \section*{INTRODUCTION}

Data on various objects are used to determine the Galactic
rotation and spiral structure parameters. Classical Cepheids,
whose distances are determined from the period.luminosity
relation, are the most important objects for solving this problem
(Feast and Whitelock 1997; Mishurov and Zenina 1999; Mel'nik et
al. 1999; Rastorguev et al. 1999).

Cepheids are distributed in a fairly wide solar neighborhood
(r$\approx$5 kpc). At present, highly accurate data are available
for quite a few of them (about 200) to determine their space
velocities.

An important parameter of the Galactic spiral structure is the
spiral pattern speed $\Omega_p$. Mishurov et al. (1979) proposed a
method for estimating this quantity using the correction factors
$F^{(1)}_\nu(x)$ and $F^{(2)}_\nu(x)$ and found
$\Omega_p=19.1\pm3.6$ km s$^{-1}$ kpc$^{-1}$ from Cepheids. Popova
(2006) obtained an estimate of $\Omega_p=21.7\pm2.8$ km s$^{-1}$
kpc$^{-1}$ also from Cepheids but by a different method based on
analysis of the ``$\ln R-\theta$'' plane. At present, however, we
have no firm confidence in the accuracy of $\Omega_p$. For
example, according to the review by Gerhard (2011), the values of
$\Omega_p$ obtained by different authors lie within the range from
15 to 30 km s$^{-1}$ kpc$^{-1}$.

The fact that our analysis of the radial velocities ($V_R$) for
Galactic masers (Bobylev and Bajkova 2010; Stepanishchev and
Bobylev 2011) and OB3 stars (Bobylev and Bajkova 2011) showed a
significant difference between these two samples of young objects
at the Sun's phase $\chi_\odot$  of about $50^\circ$ served as one
of the incentives to perform this work. This suggests that this
difference is due to the observed difference
$(\Omega_p-\Omega)\cdot t.$  If accurate estimates of the
individual or group ages for stars were available, then
information about $\Omega_p$ could be directly extracted from the
observed difference of the $\chi_\odot$ values. Cepheids are such
stars with well-known age estimates.

The goal of this paper is to determine the Galactic rotation
parameters and Galactic spiral density wave  parameters from the
space velocities of Cepheids. To estimate the spiral pattern
speed, we suggest using a direct method based on analysis of the
change in the Sun's phase with time. For this purpose, we produce
and investigate three samples of Cepheids with different mean
ages.

\section{DATA}

We used data on $\approx$240 classical Cepheids with proper
motions mainly from the Hipparcos catalog (van Leeuwen 2007) and
line-of-sight velocities from various sources. The data from
Mishurov et al. (1997) and Gontcharov (2006) as well as from the
SIMBAD and DDO databases served as the main sources of
line-of-sight velocities for the Cepheids. For several long-period
Cepheids, we used their proper motions from the TRC (Hog et al.
2000) and UCAC3 (Zacharias et al. 2009) catalogs.

To calculate the Cepheid distances, we use the calibration from
Fouqu et al. (2007). The calibration of the period.luminosity
relation for Cepheids from Berdnikov et al. (2000) is also well
known. Only nine Galactic Cepheids were used for the calibration
of Berdnikov et al. (2000). However, all of them are members of
open clusters and pulsate in the same mode. The distance
calibration for these open star clusters was first thoroughly
studied. Fouqu et al. (2007) refined the period.luminosity
relation for Cepheids using 59 calibration stars with their
parallaxes measured by various methods. According to Fouqu et al.
(2007), $M_V=-1.275-2.678\cdot\lg P,$, where the period is in
days.1 Given $M_V,$, taking the period-averaged apparent
magnitudes <$V$> and extinction $A_V=3.23\cdot E(B-V)$  mainly
from Acharova et al. (2012) and, for several stars, from Feast and
Whitelock (1997), we determine the distance $r$ from the relation
 \begin{equation}\displaystyle
 r=10^{\displaystyle -0.2(M_V-V-5+A_V)},
 \label{Ceph-02}
 \end{equation}
and then assume the relative error in the Cepheid distances
determined by this method to be 10\%. The list of Cepheids (their
numbers according to the Hipparcos catalog) from Feast and
Whitelock (1997) included in our sample is given in Table 1. The
distances from Fouqu et al. (2007) and Berdnikov et al. (2000) are
compared in Fig. 1. We see from this figure that the differences
between them are insignificant, although the scale from Berdnikov
et al. (2000) is slightly shorter. For checking, we determined all
of the sought-for parameters using the distances $r$ from
Berdnikov et al. (2000) but found no significant differences in
the parameters being determined.

\begin{table}[t]
\caption[]{\small
Cepheids from the list by Feast and Whitelock (1997) included in
our sample}
\begin{center}
      \label{t:fist}
\begin{tabular}{|r|r|c|}\hline

   HIP &    HIP  \\\hline
  5138 &  78797 \\
  5846 &  78978 \\
 34527 &  87072 \\
 39144 &  89013 \\
 42492 &  91366 \\
 42926 &  91613 \\
 50615 &  91738 \\
 51653 &  93399 \\
 52380 &  98376 \\
 53867 & 104002 \\
 56991 & 106754 \\
 57649 & 116556 \\\hline
\end{tabular}\end{center}
\end{table}

We divided the entire sample into three parts, depending on the
pulsation period, which reflects well the mean Cepheid age.
Several calibrations proposed to estimate the mean Cepheid age are
known. We use two of them: first, the theoretical calibration from
Bono et al. (2005),
\begin{equation}
 \log t_1=8.31-0.67\log P,
\label{AGE-BONO}
 \end{equation}
for the fundamental period of Cepheids with a mean metallicity of
$0.02$ typical of Galactic stars; and, second, the calibration
from Efremov (2003),
\begin{equation}
 \log t_2=8.50-0.65\log P,
\label{AGE-EFREM}
\end{equation}
obtained by analyzing Cepheids in the Large Magellanic Cloud.

We rejected the double Cepheids and short-period Cepheids
classified as DCEPs, which pulsate in the first overtone and,
therefore, their distances have a low accuracy. The main errors in
the space velocities of Cepheids are associated with the errors in
their proper motions. This is especially clearly seen in the
velocity components W. Therefore, we used a constraint on the
absolute value of the residual (after the subtraction of the
Galactic rotation parameters) velocity, $|V_{UVW}|<50$ km
s$^{-1}$. All of the remaining Cepheids are no father than 5 kpc
from the Sun.

\begin{figure}[t]
 {\begin{center}
\includegraphics[width=80mm]{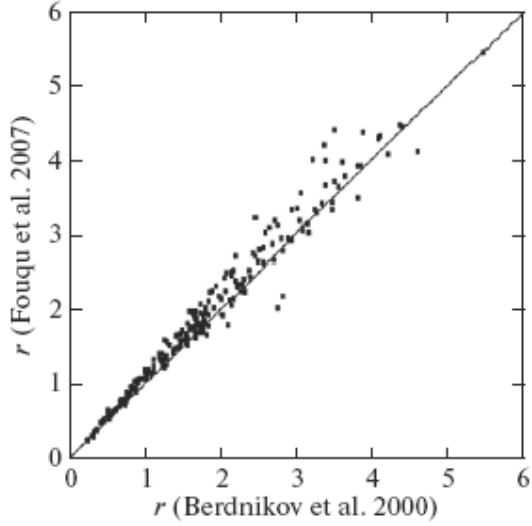}
 \caption{Cepheid distances found from the calibrations of
Berdnikov et al. (2000) and Fouqu et al. (2007).}
 \end{center}}
 \end{figure}

\section{THE METHODS}
\subsection{Simultaneous Solution}

 This method for determining the kinematic
parameters consists in minimizing a quadratic functional $F$:
 \renewcommand{\arraystretch}{1.2}
  \begin{equation}
   \begin{array}{rll}
  \min~F&=&\sum_{j=1}^N w_r^j (V_r^j-\hat{V}_{r}^j)^2\\
        &+&\sum_{j=1}^N w_l^j (V_l^j-\hat{V}_{l}^j)^2
  \label{Functional}
 \end{array}
\end{equation}
provided the fulfilment of the following constraints derived from
Bottlinger's formulas with an expansion of the angular velocity of
Galactic rotation $\Omega$ into a series to terms of the second
order of smallness with respect to $r/R_0,$ and with allowance
made for the influence of the spiral density wave:
 \begin{equation}
 \begin{array}{rll}
 V_r&=&-u_\odot\cos b\cos l\\
    &-&v_\odot\cos b\sin l-w_\odot\sin b\\
 &+&R_0(R-R_0)\sin l\cos b \Omega'_0\\
 &+&0.5R_0 (R-R_0)^2 \sin l\cos b \Omega''_0\\
 &+&\tilde{v}_\theta\sin(l+\theta)\cos b,\\
 &-&\tilde{v}_R \cos(l+\theta)\cos b,
 \label{EQ-1}
 \end{array}
 \end{equation}

 \begin{equation}
 \begin{array}{rll}
 V_l&=& u_\odot\sin l-v_\odot\cos l\\
  &+&(R-R_0)(R_0\cos l-r\cos b) \Omega'_0\\
  &+&(R-R_0)^2 (R_0\cos l - r\cos b)\times\\
&\times&0.5\Omega''_0 - r \Omega_0 \cos b\\
  &+&\tilde{v}_\theta \cos(l+\theta)+\tilde{v}_R\sin(l+\theta),
 \label{EQ-2}
 \end{array}
 \end{equation}
where $N$ is the number of stars used; $j$ is the current star
number; $V_r$ is the line-of-sight velocity, $V_l=4.74 r \mu_l\cos
b$ and $V_b=4.74 r \mu_b$ are the proper motion velocity
components in the $l$ and $b$ directions, respectively, with the
coefficient 4.74 being the quotient of the number of kilometers in
an astronomical unit and the number of seconds in a tropical year;
$\hat{V}_{r}^j, \hat{V}_{l}^j$ are the measured components of the
velocity field (data); $w_r^j$ and $w_l^j$ are the weight factors;
$r$ is the star's heliocentric distance; the star's proper motion
components $\mu_l\cos b$ and $\mu_l\cos b$ and $\mu_b$ are in mas
yr$^{-1}$ and the line-of-sight velocity $V_r$ is in km s$^{-1}$;
$u_\odot,v_\odot,w_\odot$ are the stellar group velocity
components relative to the Sun taken with the opposite sign (the
velocity $u$ is directed toward the Galactic center, $v$ is in the
direction of Galactic rotation, $w$ is directed to the north
Galactic pole), we assume $w_\odot$ to be 7 km s$^{-1}$, because
it is poorly determined without invoking the velocity components
$V_b$; $R_0$ is the Galactocentric distance of the Sun; $R$ is the
distance from the star to the Galactic rotation axis,
\begin{equation}
  R^2=r^2\cos^2 b-2R_0 r\cos b\cos l+R^2_0.
 \label{RR}
\end{equation}
$\Omega_0$ is the angular velocity of rotation at the distance
$R_0$; the parameters $\Omega'_0$ and $\Omega''_0$ are,
respectively, the first and second derivatives of the angular
velocity. To take into account the influence of the spiral density
wave, we used the simplest kinematic model based on the linear
density wave theory by Lin and Shu (1964), in which the potential
perturbation is in the form of a travelling wave. Then
\begin{equation}
 \begin{array}{rll}
      \tilde{v}_R&=&f_R \cos \chi,\\
 \tilde{v}_\theta&=&f_\theta \sin \chi,
 \label{VR-Vtheta}
 \end{array}
 \end{equation}
where $f_R$ and $f_\theta$ are the amplitudes of the radial
(directed toward the Galactic center in the arm) and azimuthal
(directed along the Galactic rotation) velocity perturbations; the
wave phase $\chi$ in general form is (Rohlfs 1977)
\begin{equation}
   \chi=m[\Omega_p-\Omega(R)] t+\Phi(R),
 \label{chi-tt}
 \end{equation}
where for a logarithmic spiral $\Phi(R)=\cot (i)\ln (R/R_0)$, Eq.
(9) for a fixed instant of time $t$ can be written as
 \begin{equation}
   \chi=m[\cot (i)\ln (R/R_0)-\theta]+\chi_\odot,
 \label{chi-creze}
 \end{equation}
$f_R$ and $f_\theta$  enter into Eqs. (5) and (6) precisely for
this approximation; $i$ is the spiral pitch angle ($i<0$ for
winding spirals); $m$ is the number of arms, we take $m=2$ in this
paper;  $\theta$ is the star's position angle (measured in the
direction of Galactic rotation); $\chi_\odot$ is the Sun's phase
angle, measured here from the center of the Carina-Sagittarius
spiral arm (R . 7 kpc), as was done by Rohlfs (1977). The
parameter $\lambda$ is the distance (along the Galactocentric
radial direction) between adjacent segments of the spiral arms in
the solar neighborhood (the wavelength of the spiral density
wave).it is calculated from the relation
 \begin{equation}
 \tan i = {{\lambda m}\over {2\pi R_0}}.
 \label{tan-i}
 \end{equation}
The weight factors in functional (4) are assigned according to the
following expressions (for simplification, we omit the index j):
 \begin{equation}
 \begin{array}{rll}
 w_r&=&S_0/\sqrt {S_0^2+\sigma^2_{V_r}},\\
 w_l&=&\beta^2 S_0/\sqrt {S_0^2+\sigma^2_{V_l}},
 \label{WESA}
 \end{array}
 \end{equation}
where $S_0$ denotes the dispersion averaged over all observations,
which has the meaning of a ``cosmic'' dispersion taken to be 12 km
s$^{-1}$; $\beta=\sigma_{V_r}/\sigma_{V_l}=1$ is the scale factor
that we determined using data on open star clusters (Bobylev et
al. 2007). The errors of the velocities  $V_l$ are calculated from
the formula
 \begin{equation}
 \sigma_{V_l}=4.74r\sqrt{\mu^2_{l}\Biggl({\sigma_r\over r}\Biggr)^2+\sigma^2_{\mu_l}}.
 \label{Errors}
 \end{equation}
The optimization problem (4).(12) is solved for nine unknown
parameters  $u_\odot,$ $v_\odot,$ 
 $\Omega_0$, $\Omega'_0,$ $\Omega''_0,$ $f_R,$
 $f_\theta,$ $i$ and $\chi_\odot$   by the
coordinate-wise descent method.

We estimated the errors of the sought-for parameters through Monte
Carlo simulations. The errors were estimated by performing 1000
cycles of computations. For this number of cycles, the mean values
of the solutions essentially coincide with the solutions obtained
from the input data without any addition of measurement errors.
Measurements errors were added to such input data as the
line-of-sight velocities, proper motions, and distances.

Here, we take a fixed value of R0 based on the review by Foster
and Cooper (2010), where the weighted mean was $R_0=8.0\pm0.4$
kpc.

Once the parameters of the rotation curve $\Omega(R),$ the
perturbation amplitudes $f_R$ and $f_\theta$, and the pitch angle
$i$ have been found from the solution of the system of equations
(5), (6), we are able to determine $\Omega_p$. For this purpose,
we use the approach applied by Mishurov et al. (1979). It is based
on the assumption about an ellipsoidal distribution of residual
stellar velocities. According to Lin et al. (1969),
 \begin{equation}
       f_R = {k A\over\kappa} {\nu\over 1-\nu^2} F^{(1)}_\nu(x),
 \label{Factors-1}
 \end{equation}

 \begin{equation}
  f_\theta=-{k A\over2\Omega} {1\over 1-\nu^2} F^{(2)}_\nu(x),
 \label{Factors-2}
 \end{equation}
where $A$ is the amplitude of the spiral density wave potential,
$\kappa^2=4\Omega^2\left(1+{\displaystyle R\over\displaystyle
2\Omega}
 {\displaystyle d\Omega\over\displaystyle dR}\right)$
 is the epicyclic frequency,
$\nu=m(\Omega_p-\Omega)/\kappa$ is the frequency with which a test
particle encounters the passing spiral perturbation,
$k=m\cdot\cot(i)/R_0$ is the radial wave number, $F^{(1)}_\nu(x)$
and $F^{(2)}_\nu(x)$ are the reduction factors, which are
functions of the coordinate $x=k^2\sigma^2_R/\kappa^2,$, where
 $\sigma_R$ is the semimajor axis of the velocity ellipsoid. An expanded
form of Eqs. (14) and (15) can be found in Mishurov et al. (1979).

Once Eqs. (14), (15) have been solved, we are able to estimate the
ratio of the radial component of the gravitational force produced
by the spiral arms to the total gravitational force of the Galaxy,
$f_{r0},$, based on the well-known relation (Fernandez et al.
2008; Bobylev et al. 2011)
 \begin{equation}
 A=\frac{(R_0\Omega_0)^2 f_{r0} \tan i}{m}.
 \label{f-r0}
 \end{equation}
Our proposed approach to estimating the spiral pattern speed
consists in finding the shifts in phase $\chi_\odot$ from several
samples of stars with an age difference $\Delta t$  and to
determine $\Omega_p$ from relation (9) for the known $\Omega(R_0)$
 \begin{equation}
   \Delta \Omega=\Omega_p-\Omega(R_0)= {\Delta \chi_\odot\cdot 1000\over m \Delta t},
 \label{MY-chi}
 \end{equation}
where the phase difference $\Delta \chi_\odot$ is in radians, and
the age difference $\Delta t$ is in Myr. Note that the change in
the Sun's phase does not depend on the Galactic orbit of the Sun;
we just find all of the parameters being determined for $R=R_0$
and, therefore, it would be more appropriate to call $\chi_\odot$
the phase of the solar circle.

\subsection{Separate Approach}

In the separate approach, a periodogram analysis based on the
Fourier transform to determine the periodicities in the velocity
components $V_R$  and $\Delta V_{rot}$  is used (Bobylev et al.
2008; Bobylev and Bajkova 2010). In this paper, we apply the
method of Fourier analysis that takes into account the logarithmic
pattern of the spiral density wave and the distribution of
position angles $\theta$ (Bajkova and Bobylev 2012). .

Initially, we define the heliocentric components of the Cepheid
space velocities $U$ and $V$ via the observed velocities
$V_r,V_l,$ and $V_b$ based on the well-known relations
 \begin{equation}
 \begin{array}{rll}
 U&=&V_r\cos l\cos b-V_l\sin l-V_b\cos l\sin b,\\
 V&=&V_r\sin l\cos b+V_l\cos l-V_b\sin l\sin b,
 \label{UVW}
 \end{array}
 \end{equation}

Subsequently, we find two velocity components: the radial velocity
$V_R$ directed from the Galactic center to the object and the
tangential velocity $V_{\theta}$ in the direction of Galactic
rotation:
 \begin{equation}
 \begin{array}{rll}
 V_{\theta}&=& U\sin \theta+(V_0+V)\cos \theta, \\
        V_R&=&-U\cos \theta+(V_0+V)\sin \theta,
 \label{U-R-T}
 \end{array}
 \end{equation}
where $V_0=|R_0\Omega_0|$ and the position angle $\theta$ is
calculated as $\tan\theta=y/(R_0-x)$, where $x$ and $y$ are the
heliocentric rectangular coordinates of the stars. Finally, based
on the velocities $V_\theta$, we form the residual tangential
velocities of the Cepheids $\Delta V_{rot}$ from which the
rotation curve found above was subtracted.

In contrast to the previous method, here we do not assume that the
wavelength $\lambda$  is the same for the velocity perturbations
in the radial and tangential directions and that their phases
$\chi_\odot$ differ exactly by $\pi/2$.

Below, we describe the method of spectral analysis of the radial
velocity perturbations for objects both in the linear
approximation of the dependence of the argument in (10) on
Galactocentric distance and in its exact expression, i.e., by
taking into account the logarithmic dependence on Galactocentric
distance and position angle $\theta$.

Let there be a series of measured velocities $V_{R_n}$  (these can
be both radial, $V_R,$, and tangential, $\Delta V_{rot}$,
velocities), $n=1,\dots,N$, where $N$ is the number of objects.
The goal of the spectral analysis is to separate the periodicity
from the data series in accordance with model (8)-(11), which
describes a spiral density wave with parameters $f_R, \lambda
(i)$, and $\chi_\odot$.

Given relation (11) between the spiral pitch angle $i$ and
wavelength $\lambda$ for $|R-R_{\circ}|\ll R_{\circ}$ and small
$\theta$, the linear approximation for the logarithm of the
argument in (10) can be represented as
$$
\frac {2\pi R_{\circ}}{\lambda} \ln(R/R_{\circ})\approx \frac
{2\pi (R-R_{\circ})}{\lambda}.
$$
In this case, for our harmonic analysis of the velocities, we can
apply the standard Fourier transform
\begin{equation}
 \bar{V}_{\lambda_k} = \frac{1} {N}\sum_{n=1}^{N} V_{R_n}
 \exp\Bigl(-j\frac{2\pi}{\lambda_k}(R_n-R_{\circ})\Bigr),
 \label{19}
\end{equation}
where $\bar{V}_{\lambda_k}$ is the $k$th harmonic of the Fourier
transform, $V_{R_n}$ are the velocity measurements for objects
with Galactocentric distances $R_n,n=1,2,...,N$, and $\lambda_k$
is the wavelength of the $k$th harmonic, which is equal to $D/k$,
where $D$ is the period of the series being analyzed.

Since we are interested only in the perturbation power spectrum
(periodogram) $|\bar{V}_{\lambda_k}|^2$ |, Eq. (20) can be
simplified as follows:
$$
 \bar{V}_{\lambda_k} = \frac{1} {N}\sum_{n=1}^{N} V_{R_n}
 \exp\Bigl(-j\frac{2\pi}{\lambda_k}R_n\Bigr).
 \label{20}
$$
Note that the derived linear approximation is acceptable only for
analyzing the perturbations in a small solar neighborhood, while
an exact realization of relation (10) is required for a wide
scatter of distances and position angles for the objects.

Let us analyze the perturbations as a periodic function of the
logarithm of the Galactocentric distances, for the time being,
without allowance for the position angles of the objects:
\begin{equation}
 \bar{V}_{\lambda_k}=\frac{1} {N}\sum_{n=1}^{N} V_{R_n}
 \exp\Bigl(-j\frac {2\pi R_{\circ}}{\lambda_k}\ln(R_n/R_{\circ})\Bigr).
\end{equation}
Obviously, if we make the change of variables
\begin{equation}
 R^{'}_{n}=\ln(R_n/R_{\circ})R_{\circ},
 \label{21}
\end{equation}
is reduced to the standard Fourier transform
\begin{equation}
 \bar{V}_{\lambda_k} = \frac{1} {N}\sum_{n=1}^{N} V_{R^{'}_n}
 \exp\Bigl(-j\frac {2\pi R^{'}_n}{\lambda_k}\Bigr).
 \label{22}
\end{equation}
To take the position angles of the objects into account, we will
represent Eq. (10) for the phase as
\begin{equation}
 \chi = \chi_1-m\theta,
 \label{23}
\end{equation}
where (given (11))
\begin{equation}
 \chi_1 = \frac {2\pi R_{\circ}}{\lambda}\ln(R/R_{\circ})+\chi_{\odot}.
 \label{24}
\end{equation}
Substituting (24) into Eq. (10) for the perturbations at the $n$th
point and making standard trigonometric transformations, we will
obtain:
 \begin{equation}
 \begin{array}{rll}
 &V_{R_n}=f_R \cos(\chi_{1_n} - m\theta_n)\\
 &      =f_R\cos\chi_{1_n}\cos m\theta_n   + f_R\sin\chi_{1_n}\sin m\theta_n \\
 &      =f_R \cos\chi_{1_n}(\cos m\theta_n + \tan\chi_{1_n}\sin  m\theta_n).
 \label{25}
 \end{array}
 \end{equation}
Let us designate
\begin{equation}
 V_{R^{'}}=f_R\cos\chi_1,
 \label{26}
\end{equation}
It then follows from (26) that
\begin{equation}
 V_{R_n}=V_{R^{'}_n}(\cos m\theta_n+\tan\chi_{1_n} \sin m\theta_n).
 \label{27}
 \end{equation}
 Using Eq. (28), let us form a new data series
\begin{equation}
 V_{R^{'}_n}=V_{R_n}/(\cos m\theta_n+\tan\chi_{1_n} \sin m\theta_n),
 \label{28}
\end{equation}
to which a Fourier analysis can be applied in accordance with
(23). A similar relation can also be derived for the tangential
velocities $\Delta V_{rot}$.

Thus, taking into account both the logarithmic pattern of the
spiral density wave and the position angles of the objects, we
obtain the following expression for our spectral analysis of the
perturbations:
\begin{equation}
 \bar{V}_{\lambda_k} = \frac{1} {N}\sum_{n=1}^{N} V^{'}_{R^{'}_n}
 \exp\Bigl(-j\frac {2\pi R^{'}_n}{\lambda_k}\Bigr).
 \label{29}
\end{equation}

The numerical algorithm for realizing (30) consists of the
following steps:

1. The initial series of velocities $V_{R_n}$ is transformed into
the series $V_{R_n^{'}}$ in accordance with (22).

2. The power spectrum of the derived sequence $V_{R_n^{'}}$ is
calculated based on the Fourier transform (23) to obtain an
estimate of $\lambda_{max}$ that corresponds to the peak of the
derived power spectrum.

3. A comb of several $\lambda_i (i=1,\dots,K)$  with a central
$\lambda_{max}$ is then specified.

The following iterations are made for each $\lambda_i$ from the
specified comb:

Step 1. The value of $\lambda_i$ and the initial approximation
$\chi_\odot$ (for example, equal to zero) are substituted into Eq.
(25) to calculate $\chi_1$ for each data reading ($n=1,\dots,N$).

Step 2. Using Eq. (29), the series of velocities $V_{R_n^{'}}$  is
transformed into the series $V^{'}_{R^{'}_n}$. This transformation
needs to be regularized to avoid the division by numbers close to
zero.This is done by assigning a threshold number, say,
$\varepsilon$, and permission for the division is given only when
the denominator in Eq. (29) exceeds this number. The best
$\varepsilon$ at which the significance of the extracted peak in
the spectrum reaches its maximum as a result of the iterations at
the minimum residual between the solution and the data can be
found by an exhaustive search for $\varepsilon$ from some
interval. The typical values of $\varepsilon$ found on model
problems lie within the range [0.01, 0.3].

Step 3. The power spectrum of the derived sequence
$V^{'}_{R^{'}_n}$ is calculated based on the Fourier transform
(30) to obtain a new estimate of $\chi_\odot$ corresponding to a
fixed $\lambda_i$  of the derived power spectrum.

Step 4. The return to the first step is made until the process
will converge or diverge.

Step 5. If the process converged, then we fix the specified
$\lambda_i$ and the derived $\chi_\odot$; if it diverged, then we
take the next value $\lambda_{i+1}$ from the specified comb and
make iterations 1--4 until the value of $\lambda$ at which the
process converges will be found.

4. The power spectrum is calculated for the values of $\lambda$
and $\chi_\odot$ found based on Eq. (30) with the goal of a
further analysis.

A more detailed description of the algorithm, the results of its
testing on model data, and the questions of estimating the
significance of the extraction of peaks in the periodogram and
estimating the errors in the spiral density wave parameters are
given in Bajkova and Bobylev (2012).

\begin{table}[t]
\caption[]{\small\baselineskip=1.0ex
  Kinematic parameters found from Cepheids
  }
\begin{center}
      \label{t:999}
\begin{tabular}{|l|c|c|c|c|c|}\hline
 Parameters                   & $P\geq 9^d$   & $5^d\leq P<9^d$ & $P<5^d$    & All         \\\hline

 $u_\odot,$    km s$^{-1}$    & $  6.9\pm1.2 $ & $ 7.5\pm0.9$ & $ 8.6\pm1.3$ & $ 7.6\pm0.5$ \\
 $v_\odot,$    km s$^{-1}$    & $ 11.6\pm0.8 $ & $10.9\pm0.6$ & $15.6\pm1.3$ & $11.7\pm0.3$ \\
 $\Omega_0,$   km s$^{-1}$ kpc$^{-1}$   & $ 26.1\pm0.9 $ & $ 30.4\pm1.0$  & $ 24.1\pm1.2$  & $ 27.5\pm0.5 $ \\
 $\Omega^{'}_0,$ km s$^{-1}$ kpc$^{-2}$ & $-3.95\pm0.13$ & $-4.34\pm0.13$ & $-4.72\pm0.25$ & $-4.12\pm0.10$ \\
$\Omega^{''}_0,$ km s$^{-1}$ kpc$^{-3}$ & $ 0.79\pm0.10$ & $ 0.69\pm0.14$ & $ 2.44\pm0.28$ & $ 0.85\pm0.07$ \\
 $f_R,$        km s$^{-1}$    & $ -9.8\pm1.3$  & $-8.5\pm1.1$   & $-12.6\pm1.7$  & $ -6.8\pm0.7$  \\
 $f_\theta,$   km s$^{-1}$    & $  1.3\pm1.9$  & $ 2.7\pm1.1$   & $ 8.2\pm1.4$   & $  3.3\pm0.5$  \\
 $i,$          deg        & $ -5.2\pm0.3$  & $-4.0\pm0.1$   & $-6.6\pm0.5$   & $ -4.6\pm0.1$  \\
 $\chi_\odot,$ deg        & $ -148\pm14$   & $-193\pm9$     & $-234\pm10$    & $ -193\pm5$    \\
 $\sigma_0,$   km s$^{-1}$    &          12.2  &         12.5   &         12.5   &          13.4  \\
 $\lambda,$    kpc            & $  2.3\pm0.5$  & $ 1.8\pm0.1$   & $ 2.9\pm0.4$   & $  2.0\pm0.1$  \\
 $N_\star$                    &            61  &           72   &           52   &           185  \\
 $\overline t_1$, Myr    &          33.9  &         59.2   &         84.9   &                \\
 $\overline t_2$, Myr    &          55.4  &         95.2   &        135.0   &                \\
 $\sigma_R,$      km s$^{-1}$ &                &                &                &           14   \\
 $\Omega_p$, km s$^{-1}$ kpc$^{-1}$ &                &                &                &         23.5   \\
 $f_{r0}$       &                &                &                & $0.04\pm0.01$ \\\hline

\end{tabular}
\end{center}
{\small
 Note:
 $N_\star$~is the number of Cepheids in the sample,
 $\sigma_0$~is the error per unit weight obtained when solving the system of equations (5),(6),
 $\sigma_R$~is the semimajor axis of the velocity ellipsoid,
$\Omega_p$ was estimated by the method of Mishurov et al. (1979).}
\end{table}

\section{RESULTS AND DISCUSSION}
\subsection{Results of the Simultaneous Solution}

The results of the solutions of the system of equations (5) and
(6) obtained from four samples of Cepheids are presented in Table
2: for Cepheids with periods longer than 9 days, i.e., the
youngest ones in our sample, in the first column; for Cepheids
with periods from 5 to 9 days, i.e., middle-aged ones, in the
second column; for Cepheids with periods shorter than 5 days,
i.e., the oldest ones, in the third column; the solution obtained
from all Cepheids is given in the fourth column.

The parameters of the Galactic rotation curve $\Omega_0,$
$\Omega^{'}_0$, and $\Omega^{''}_0,$ calculated from the sample of
middle-aged Cepheids are in good agreement with the results of
analyzing blue supergiants (Zabolotskikh et al. 2002), OB
associations (Mel'nik and Dambis 2009), and Galactic masers
(Bobylev and Bajkova 2010; Stepanishchev and Bobylev 2011). In
these papers, the angular velocity of Galactic rotation $\Omega_0$
is $\approx$30 km s$^{-1}$ kpc$^{-1}$. For unknown reasons, the
youngest Cepheids revolve around the Galactic center with a lower
velocity. Note that $\Omega_0$ and $\Omega^{'}_0,$ that we found
based on the entire sample of Cepheids (the last column in Table
2) are in good agreement with the results of the analysis of only
the proper motions of Cepheids from the Hipparcos catalog
performed by Feast and Whitelock (1997), where
$\Omega_0=27.2\pm0.9$km s$^{-1}$ kpc$^{-1}$ was found.

It can be seen from Table 2 that the amplitudes of the velocity
perturbations in the radial direction, $f_R,$, differ
significantly from zero in all cases, while in the tangential
direction, $f_\theta,$, they are significant for the sample of old
Cepheids (and for the entire sample). The relationship between the
amplitudes found is in agreement with the results of the analysis
of blue supergiants performed by Zabolotskikh et al. (2002),
$f_R=-6.6\pm2.5$ km s$^{-1}$ and $f_\theta=0.4\pm2.3$ km s$^{-1}$
for $m=2$, and a sample of young Cepheids with similar values of
these parameters. Previously (Bobylev and Bajkova 2011), we found
$f_R=-12.5\pm1.1$ km s$^{-1}$, $f_\theta= 2.0\pm1.6$ km s$^{-1}$,
and $i=-5.3\pm0.3^\circ$ for $m=2$ with $\chi_\odot=-91\pm4^\circ$
from data on OB3 stars with an independent distance scale
determined from interstellar Ca~II absorption lines.

In principle, the Sun's phases found (Table 2) are consistent with
the results of analyzing various samples of Cepheids, for example
(with the phase measured from the Carina-Sagittarius arm):
$\chi_\odot=-165\pm1^\circ$ (Byl and Ovenden 1978),
$\chi_\odot=-150^\circ$ from red supergiants and Cepheids
(Mishurov et al. 1979), $\chi_\odot=-290\pm16^\circ$ (Mishurov et
al. 1997), and $\chi_\odot=-320\pm9^\circ$ (Mishurov and Zenina
1999) from relatively old Cepheids with periods $P<9^d$.

In Figs. 2 and 3, the radial, $V_R$, and tangential,
$V_{rot}=V_\theta$, velocities (calculated from Eqs. (19)) are
plotted against the Galactocentric distance R for two samples of
Cepheids.young and middle-aged ones. In each case, the rotation
curve was constructed in accordance with the results of Table 2. A
wave structure in the radial velocities $V_R$ is clearly seen in
Figs.2 and 3.

 \begin{figure}[t]
 {\begin{center}
 \includegraphics[width=80mm]{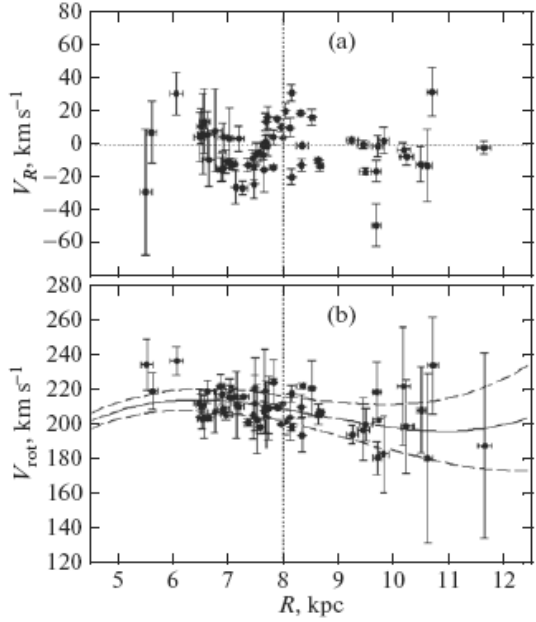}
 \caption{Radial velocities for the sample of young Cepheids
(a) and their tangential velocities (b) versus Galactocentric
distance R; the vertical dotted lines mark the Sun’s position, the
solid curve in panel (b) indicates the rotation curve, the lines
with long dashes mark the $1\sigma$ confidence intervals. }
 \label{f-young}
 \end{center} }
 \end{figure}

 \begin{figure}[t]
 {\begin{center}
 \includegraphics[width=80mm]{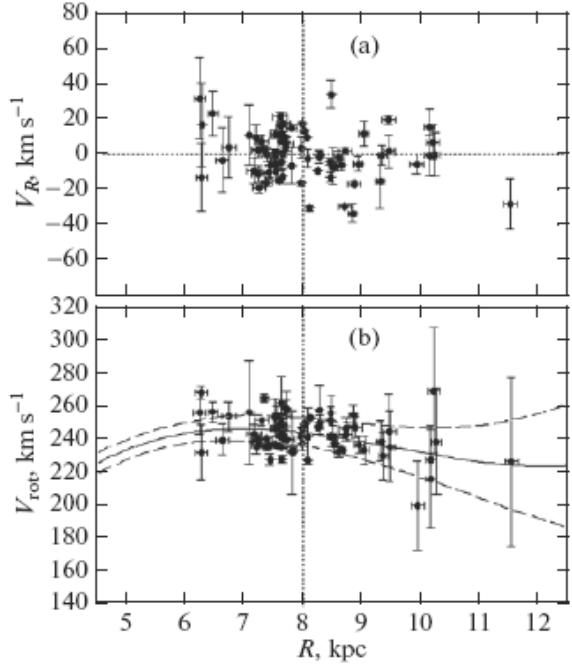}
 \caption{Radial velocities for the sample of middle-aged
Cepheids (a) and their tangential velocities (b) versus
Galactocentric distance R. The designations are the same as those
in Fig. 2.}
 \label{f-median}
 \end{center} }
 \end{figure}

As can be seen from Table 2, the phase $\chi_\odot$  changes
noticeably with the mean age of the sample. We also see that the
mean Cepheid ages calculated using calibrations (2) and (3) differ
by a factor of 1.5. Applying Eq. (17) gives
$|\Delta\Omega|_{(1-2)}= 9.9$~km s$^{-1}$ kpc$^{-1}$ (columns 1,
2, 3 in Table 2),
 $|\Delta\Omega|_{(2-3)}= 9.4$~km s$^{-1}$ kpc$^{-1}$ and
 $|\Delta\Omega|_{(1-3)}= 9.0$~km s$^{-1}$ kpc$^{-1}$
for the calibration of Efremov (2003) (Eq. (3)). In this case, the
mean value of the difference  $|\Delta\Omega|=9.4$~km s$^{-1}$
kpc$^{-1}$, then $\Omega_p=18.1$~km s$^{-1}$ kpc$^{-1}$ (for
$\Omega_0=27.5\pm0.5$~km s$^{-1}$ kpc$^{-1}$ found from the entire
sample of Cepheids), which is in satisfactory agreement with the
result obtained from Eqs. (14) and (15) by the method of Mishurov
et al. (1979), $\Omega_p=23.5$~km s$^{-1}$ kpc$^{-1}$.

The analogous values found from the age calibration of Bono et al.
(2005) (Eq. (2)) are considerably larger:
$|\Delta\Omega|_{(1-2)}=15.4$~km s$^{-1}$ kpc$^{-1}$,
 $|\Delta\Omega|_{(2-3)}=14.7$~km s$^{-1}$ kpc$^{-1}$ and
 $|\Delta\Omega|_{(1-3)}=14.0$~km s$^{-1}$ kpc$^{-1}$.
In this case, the mean value of the difference
$|\Delta\Omega|=14.7$~km s$^{-1}$ kpc$^{-1}$, then
$\Omega_p=12.8$~km s$^{-1}$ kpc$^{-1}$, which is in much poorer
agreement with other known data.

Comparison of the results obtained allows us to opt for the age
calibration of Efremov (2003).

The estimate of $f_{r0}=0.04\pm0.01$ that we obtained from Eq.
(16) is much smaller than $f_{r0}=0.15,$ found from a sample of
192 long-period Cepheids by Mishurov et al. (1979). At the same
time, our estimate is in good agreement with $f_{r0}=0.05,$
suggested by Yuan (1969).
 \begin{figure}[p]
 {\begin{center}
 \includegraphics[width=150mm]{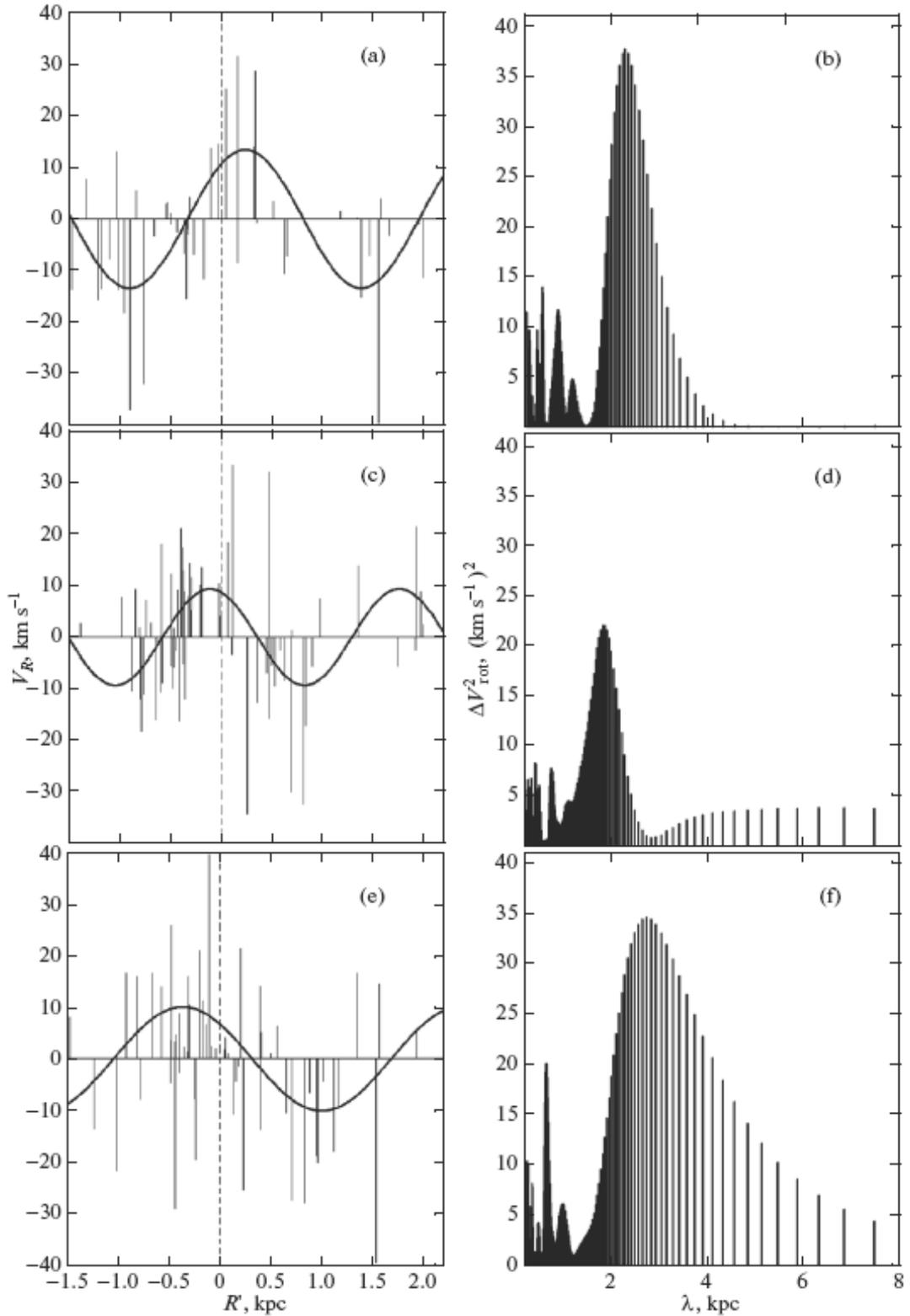}
 \caption{Radial velocities for the sample of young Cepheids (a) and their power spectrum (b), middle-aged Cepheids (c) and
their power spectrum (d), and old Cepheids (e) and their power
spectrum (f); the vertical dotted lines mark the Sun’s position.}
 \label{rad}
 \end{center} }
 \end{figure}

 \begin{figure}[t]
 {\begin{center}
 \includegraphics[width=150mm]{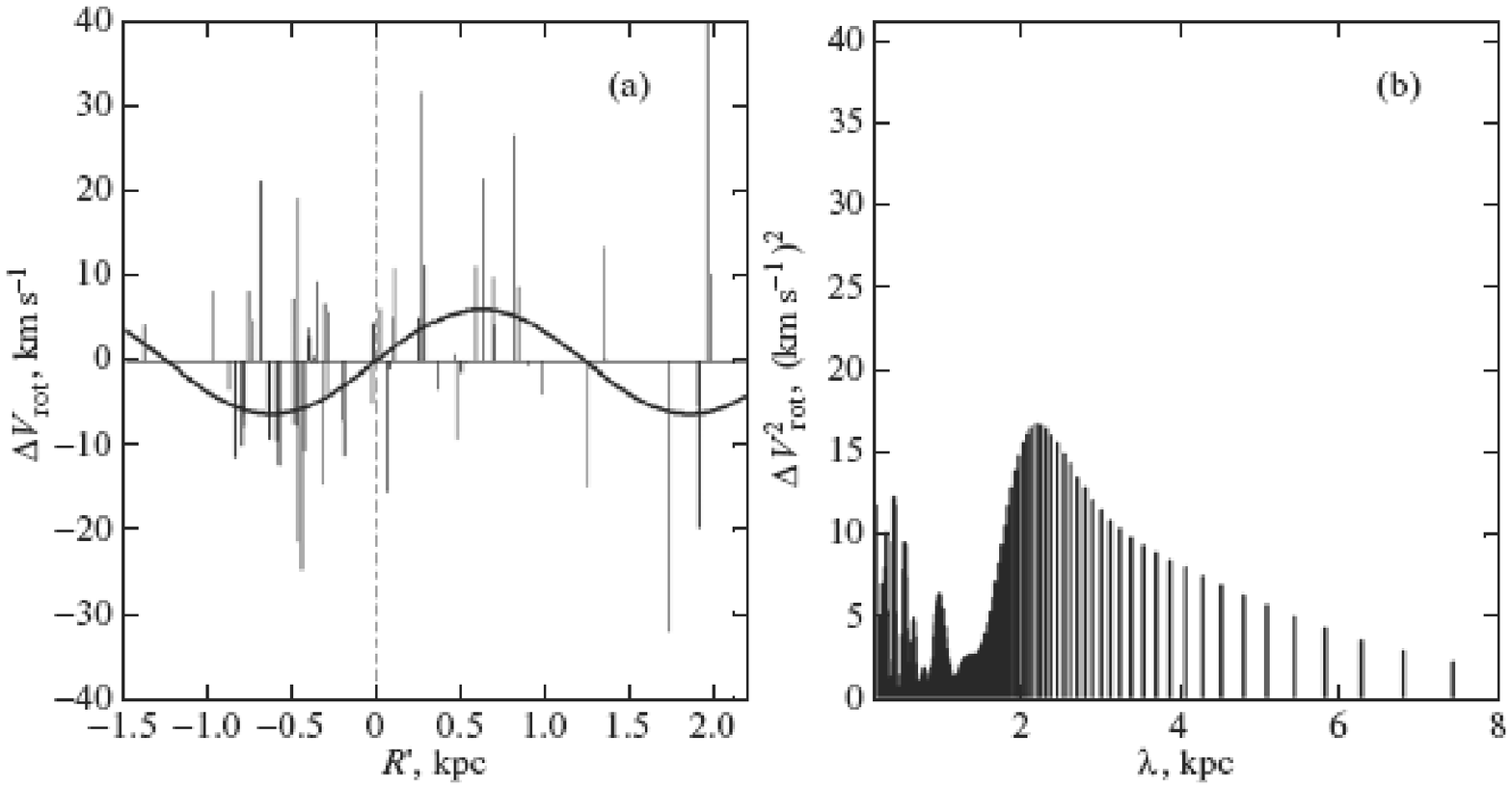}
 \caption{Residual tangential velocities for the sample of middle-aged Cepheids (a) and their power spectrum (b); the
corresponding radial velocities are shown in panels (c and d) of
Fig. 4.}
 \label{tang}
 \end{center} }
 \end{figure}

\subsection{Results of the Separate Approach}

Figure 4 presents the Galactocentric radial velocities $V_R$ of
the Cepheids and the corresponding power spectra. The distance
$R^{'}$ indicated in the plots was calculated from the relation
$R^{'}=\ln(R/R_{\circ})R_{\circ},$ which takes into account the
logarithmic pattern of the spiral density wave (22). In contrast
to Figs. 2 and. 3, the data here are presented in the form of
``pulses'' without indicating their random errors.

As in the previous case (Table 2), this approach failed to reveal
a significant wave in the tangential  velocities for the sample of
the youngest Cepheids. The proposition of the density wave theory
that the perturbations in the radial direction propagate faster
than those in the tangential one can serve as an explanation of
this fact. Therefore, the youngest stars have not yet had time to
respond to the perturbations in the tangential direction, while
the situation for the older stars gradually levels off.

Figure 5 shows the residual tangential velocities $\Delta V_{rot}$
for the sample of middle-aged Cepheids and their power spectrum.

We clearly see from Figs. 4 and 5 and the data of Table 2 that the
wavelength $\lambda$  changes, depending on the sample and the
velocity character. For example, the difference in $\lambda$
calculated from the radial and tangential velocities of
middle-aged Cepheids is about 0.4 kpc. Comparison of the data in
Fig. 5 with panels (c) and (d) in Fig. 4 leads us to conclude that
there is a phase shift close to $\pi/2$  (the exact value is
$110^\circ$), which is in agreement with the prediction of the
linear density wave model. It can be seen from the sample of
middle-ages Cepheids that the Sun is equidistant from the centers
of the two nearest segments of the spiral arms.

For a more reliable determination of the mean $\Delta\Omega$, we
invoked our previously obtained data on the  kinematics of young
OB3 stars (Bobylev and Bajkova 2011), where the Sun's phase
$\chi_\odot=-91\pm4^\circ$ was obtained. For these stars, we took
a mean age of 8 Myr. We calculated the mean
$|\Delta\Omega|=9.8\pm0.6$ km s$^{-1}$ kpc$^{-1}$ using six
independent differences formed from four values of $\chi_\odot$:
$-91^\circ,$
 $-148^\circ,$
 $-193^\circ$ and
 $-234^\circ.$
We used the mean Cepheid ages calculated from Efremov's
calibration. Here, the error
$(\varepsilon_{\Delta\Omega})_{stat}=0.6$ km s$^{-1}$ kpc$^{-1}$
was calculated from the convergence of the results. We can also
estimate the systematic error of the method,
$(\varepsilon_{\Delta\Omega})_{syst},$ from the relation
\begin{equation}
   (\varepsilon^2_{\Delta \Omega})_{syst}=
   1000^2\left(\left({\varepsilon_\chi\over m \Delta t}\right)^2+
           \left(-{\Delta \chi\cdot\varepsilon_t\over m (\Delta t)^2}\right)^2 \right),
 \label{MY-eroors}
 \end{equation}
for the adopted typical values:
 $\varepsilon_\chi=10^\circ\cdot(\pi/180^\circ),$
 $\Delta_\chi=50^\circ\cdot(\pi/180^\circ),$
 $\Delta t=50$~Myr and
 $\varepsilon_t=10$~Myr, then
 $(\varepsilon_{\Delta\Omega})_{syst}=2.8$~km s$^{-1}$
kpc$^{-1}$. As a result, we have
$|\Delta\Omega|=9.8\pm0.6_{stat}\pm2.8_{syst}$ km s$^{-1}$
kpc$^{-1}$.

Note that we found $\chi_\odot=-130\pm10^\circ$ from a sample of
masers in regions of active star formation (Bobylev and Bajkova
2010). Thus, they are intermediate between the OB3 stars and young
Cepheids. Although there are no estimates of their individual
ages, there are very massive O stars as well as high-  and
low-mass proto stars among them. The paradoxical value of
$\chi_\odot=-130^\circ$ (the expected
$\chi_\odot\approx-90^\circ$) can be explained by the fact that
the kinematics of these recently formed stars probably reflects
considerably earlier stages in the Galactic motion of the regions
of active star formation.

The Galactic spiral pattern speed calculated by our proposed
method from the entire sample of Cepheids ($\Omega_0=27.5$ km
s$^{-1}$ kpc$^{-1}$) is $\Omega_p=17.7$ km s$^{-1}$ kpc$^{-1}$.
If, however, we take $\Omega_0=30$ km s$^{-1}$ kpc$^{-1}$ known
from our analysis of masers and OB3 stars, then we obtain
$\Omega_p=20.2$ km s$^{-1}$ kpc$^{-1}$. According to these data,
the co-rotation circle in the Galaxy is located at a distance from
$R=$10 to 12 kpc. These estimates are valid for a two-armed
Galactic spiral pattern ($m=2$). In the opinion of several
authors, the Galactic spiral pattern is a four-armed one ($m=4$).
In this case (see Eq. (17)), the difference we found will be
$|\Omega_p-\Omega_0|\approx5$ km s$^{-1}$ kpc$^{-1}$. Then, the
spiral pattern speed will be $\Omega_p=22.5$ km s$^{-1}$
kpc$^{-1}$ from the entire sample of Cepheids ($\Omega_0=27.5$ km
s$^{-1}$ kpc$^{-1}$) and $\Omega_p=25$ km s$^{-1}$ kpc$^{-1}$ for
$\Omega_0=30$ km s$^{-1}$ kpc$^{-1}$; accordingly, the co-rotation
circle will be still closer to the Sun.

\section{CONCLUSIONS}

We analyzed the space velocities of Galactic Cepheids to study the
peculiarities of the Galactic spiral density wave. For this
purpose, we used 185 Cepheids with proper motions mainly from the
Hipparcos catalog and line-of-sight velocities from various
sources. We divided the entire sample into three parts, depending
on the pulsation period, which reflects well the mean Cepheid age.

First, based on the entire sample of Cepheids and taking $R_0=8$
kpc, we found the components of the peculiar solar velocity
$(u_\odot,v_\odot)=(7.6,11.7)\pm(0.5,0.3)$ km s$^{-1}$, the
angular velocity of Galactic rotation $\Omega_0 = 27.5\pm0.5$ km
s$^{-1}$ kpc$^{-1}$ and its derivatives $\Omega^{'}_0 =
-4.12\pm0.10,$ km s$^{-1}$ kpc$^{-2}$ and $\Omega^{''}_0 =
0.85\pm0.07,$  km s$^{-1}$ kpc$^{-3}$, the amplitudes of the
spiral density wave $f_R=-6.8\pm0.7$ km s$^{-1}$ and $f_\theta=
3.3\pm0.5$ km s$^{-1}$, the pitch angle of a two-armed spiral
pattern $i=-4.6\pm0.1^\circ$ (then $\lambda=2.0\pm0.1$ kpc), and
the phase of the Sun in the spiral density wave
$\chi_\odot=-193\pm5^\circ$.

The Cepheids with pulsation periods $P$ from 5 to 9 days
(middle-aged) ($\Omega_0=30.4\pm1.0$ km s$^{-1}$ kpc$^{-1}$ for
them) show the fastest Galactic rotation. The amplitude of the
radial velocity perturbations $f_R$ caused by the spiral density
wave in each of the three samples is about 9 km s$^{-1}$; the
amplitude of the tangential velocity perturbations $f_\theta$
increases from zero for the youngest Cepheids to $\approx$4 km
s$^{-1}$ for older ones.

We found that the Sun's phase $\chi_\odot$ changes noticeably with
the mean age of the sample. For a more accurate estimation of the
phase, we applied a Fourier analysis of the Cepheid radial
velocities, while for middle-aged Cepheids we also managed to
reliably determine the parameters of the tangential velocity
perturbations by this method.

From our analysis of the phase shifts, we determined the mean
value of the angular velocity difference
$\Delta\Omega=\Omega_p-\Omega$, which depends significantly on the
calibrations used to estimate the individual ages of Cepheids. For
a more reliable determination of the mean $\Delta\Omega$, we
invoked our previously obtained data on the kinematics of young
OB3 stars. As a result, we calculated the mean $\Delta\Omega$
using six independent differences.

We showed that when the individual ages of Cepheids derived from
the period-age calibration of Bono et al. (2005) is used, the
differences $\Omega_p-\Omega_0$ exceed those derived from the
calibration of Efremov (2003) by a factor of 1.5-2.
Simultaneously, we obtained an estimate of $\Omega_p=23.5$ km
s$^{-1}$ kpc$^{-1}$ from our analysis of the reduction factors by
the method of Mishurov et al. (1979). As a result, this allowed us
to opt for Efremov's calibration , using which we found
$|\Omega_p-\Omega_0|=9.8\pm0.6_{stat}\pm2.8_{syst}$ km s$^{-1}$
kpc$^{-1}$.

The Galactic spiral pattern speed that we calculated by our
proposed method based on the entire sample of Cepheids
($\Omega_0=27.5$ km s$^{-1}$ kpc$^{-1}$) is $\Omega_p=17.7$ km
s$^{-1}$ kpc$^{-1}$. If, however, we take $\Omega_0=30$ km
s$^{-1}$ kpc$^{-1}$ known from our analysis of masers and OB3
stars, then we will obtain $\Omega_p=20.2$ km s$^{-1}$ kpc$^{-1}$.
According to these data, the co-rotation circle in the Galaxy is
located at a distance from R = 10 to 12 kpc.

Based on the entire sample of Cepheids, we estimated the ratio of
the radial component of the gravitational force produced by the
spiral arms to the total gravitational force of the Galaxy to be
$f_{r0}=0.04\pm0.01$.

\section{ACKNOWLEDGMENTS}

 We are grateful to the referees for valuable
remarks that contributed to a significant improvement of the
paper. This work was supported in part by the ``Nonstationary
Phenomena in Objects of the Universe'' Program of the Presidium of
the Russian Academy of Sciences and grant no. NSh-1625.2012.2 from
the President of the Russian Federation. In our work, we used the
SIMBAD search database and the DDO Cepheid database.

\section{REFERENCES }

\noindent 1. I. A. Acharova, Yu. N. Mishurov, and V. V. Kovtyukh,
Mon. Not. R. Astron. Soc. 420, 1590 (2012).

\noindent 2. A. T. Bajkova and V. V. Bobylev, Astron. Lett. 38,
549 (2012).

\noindent 3. L. N. Berdnikov, A. K. Dambis, and O. V. Vozyakova,
Astron. Astrophys. Suppl. Ser. 143, 211 (2000).

\noindent 4. V. V. Bobylev, A. T. Bajkova, S. V. Lebedeva, Astron.
Lett. 33, 720 (2007).

\noindent 5. V. V. Bobylev, A. T. Bajkova, and A. S.
Stepanishchev, Astron. Lett. 34, 515 (2008).

\noindent 6. V.V.Bobylev, andA.T. Bajkova,Mon. Not.R. Astron. Soc.
408, 1788 (2010).

\noindent 7. V. V. Bobylev and A. T. Bajkova, Astron. Lett. 37,
526 (2011).

\noindent 8. V. V. Bobylev, A. T. Bajkova, A. Myllari, and M.
Valtonen, Astron. Lett. 37, 550 (2011).

\noindent 9. G. Bono, M. Marconi, S. Cassisi, et al., Astrophys.
J. 621, 966 (2005).

\noindent 10. J. Byl and M. W. Ovenden, Astrophys. J. 225, 496
(1978).

\noindent 11. Yu. N. Efremov, Astron.Rep. 47, 1000 (2003).

\noindent 12. M. Feast and P.Whitelock, Mon. Not. R. Astron. Soc.
291, 683 (1997).

\noindent 13. D. Fernandez, F. Figueras, and J. Torra, Astron.
Astrophys. 480, 735 (2008).

\noindent 14. T. Foster and B. Cooper, ASP Conf. Ser. 438, 16
(2010).

\noindent 15. P. Fouqu, P. Arriagada, J. Storm, et al., Astron.
Astrophys. 476, 73 (2007).

\noindent 16. O. Gerhard, Mem. Soc. Astron. Ital. Suppl. 18, 185
(2011).

\noindent 17. G. A. Gontcharov, Astron. Lett. 32, 759 (2006).

\noindent 18. E. Hog, C. Fabricius, V. V. Makarov, et al., Astron.
Astrophys. 355, L27 (2000).

\noindent 19. F. van Leeuwen, Astron. Astrophys. 474, 653 (2007).

\noindent 20. C. C. Lin and F. H. Shu, Astrophys. J. 140, 646
(1964).

\noindent 21. C. C. Lin, C. Yuan, and F. H. Shu, Astrophys. J.
155, 721 (1969).

\noindent 22. A. M. Mel'nik, A. K. Dambis, and A. S. Rastorguev,
Astron. Lett. 25, 518 (1999).

\noindent 23. A. M. Mel'nik and A. K. Dambis, Mon. Not. R. Astron.
Soc. 400, 518 (2009).

\noindent 24. Yu. N. Mishurov, I. A. Zenina, A. K. Dambis, et al.,
Astron. Astrophys. 323, 775 (1997).

\noindent 25. Yu. N. Mishurov, E. D. Pavlovskaya, and A. A.
Suchkov, Sov. Astron. 23, 147 (1979).

\noindent 26. Yu. N.Mishurov and I. A. Zenina, Astron. Astrophys.
341, 81 (1999).

\noindent 27. M. E. Popova, Astron. Lett. 32, 244 (2006).

\noindent 28. A.S.Rastorguev,E. V. Glushkova, A. K. Dambis, and M.
V. Zabolotskikh, Astron. Lett. 25, 595 (1999).

\noindent 29. K. Rohlfs, Lectures on Density Wave Theory
(Springer, Berlin, 1977).

\noindent 30. A. S. Stepanishchev and V. V. Bobylev, Astron. Lett.
37, 254 (2011).

\noindent 31. C. Yuan, Astrophys. J. 158, 889 (1969).

\noindent 32. M. V. Zabolotskikh, A. S. Rastorguev, and A. K.
Dambis, Astron. Lett. 28, 454 (2002).

\noindent 33. N. Zacharias, C. Finch, T. Girard, et al., CDS
Strasbourg, I/315 (2009)

\noindent 34. The Hipparcos and Tycho Catalogues, ESA SP-1200
(1997).

\end{document}